\documentclass[onecolumn,showpacs,preprintnumbers,amsmath,amssymb]{revtex4}
\usepackage{graphicx,psfrag}% Include figure files
\usepackage{dcolumn}% Align table columns on decimal point
\usepackage{bm,bbm}% bold math

\begin{document}

\title{Two-photon interactions with Majorana fermions}

\author{David C. Latimer}

\affiliation{Department of Physics, University of Puget Sound,
Tacoma, WA 98416-1031
}

\newcommand*{\sech}{\mathop{\mathrm{sech}}\limits}
\newcommand*{\balpha}{\boldsymbol{\alpha}} 
\newcommand*{\dilog}{\mathrm{Li}_2}
\newcommand{\qslash}{\not{\hbox{\kern-2pt $q$}}}
\newcommand{\kslash}{\not{\hbox{\kern-2pt $k$}}}
\newcommand{\pslash}{\not{\hbox{\kern-2pt $p$}}}
\newcommand{\delslash}{\not{\hbox{\kern-3pt $\partial$}}}
\newcommand{\Dslash}{\not{\hbox{\kern-3pt $D$}}}
\newcommand{\gmn}{g^{\mu \nu}}
\newcommand{\Pslash}{\not{\hbox{\kern-2.3pt $P$}}}
\newcommand{\Kslash}{\not{\hbox{\kern-2.3pt $K$}}}
\newcommand{\Pslashsup}{^\not{\hbox{\kern-0.5pt $^P$}}}
\newcommand{\Poddup}{^\not{\hbox{\kern-0.5pt $^\mathcal{P}$}}}
\newcommand{\Podddown}{_\not{\hbox{\kern-0.5pt $_\mathcal{P}$}}}
\newcommand{\bsig}{\boldsymbol{\sigma}}
\newcommand{\beps}{\boldsymbol{\epsilon}}
\newcommand{\phat}{\hat{\mathbf{p}}}
\newcommand{\khat}{\hat{\mathbf{k}}}
\newcommand{\al}[1]{\begin{align}#1\end{align}}

\begin{abstract}

Because Majorana fermions are their own antiparticles, their electric and magnetic dipole moments must vanish, leaving the anapole moment as their only static electromagnetic property. But the existence of induced dipole moments is not necessarily prohibited.  Through a study real Compton scattering, we explore the constraints that the Majorana fermion's self-conjugate nature has on induced moments.  In terms of the Compton amplitude, we find no constraints if the interactions are separately invariant under charge conjugation, parity, and time reversal.  However, if the interactions are odd under parity and even under time reversal, then these contributions to the Compton amplitude must vanish.  We employ a simple model to confirm these general findings via explicit calculation of the Majorana fermion's polarizabilities.  We then use these polarizabilities to estimate the cross-section for $s$-wave annihilation of two Majorana fermions into photons.  The cross-section is larger than a na\"ive estimate might suggest.

\end{abstract}

\maketitle

\section{Introduction}

The static electromagnetic (EM) properties of spin-$\frac{1}{2}$ Majorana fermions are highly constrained.  Because these particles are their own antiparticles, they must be electrically neutral with vanishing electric and magnetic dipole moments.  In fact, their only non-zero static EM property is the anapole moment \cite{bk82,nieves,bk83,bk84}.  Within the Standard Model (SM), the neutrino is the only possible Majorana fermion.  Several decades ago, interest arose in the neutrino's EM properties, and the program of discovery is ongoing both from a theoretical and experimental standpoint \cite{deGouvea:2013zp,Chen:2014ypv}.  In terms of the neutrino's anapole moment, there was much controversy surrounding its status as a meaningful observable in part due to questions about its gauge invariance \cite{musolf,dvornikov}. Reference \cite{Giunti:2014ixa} contains a complete review of this debate; in the end, the neutrino's anapole moment was determined to be finite and gauge invariant.
With that said, our primary interest in this paper is in physics beyond the SM, where theories are rife with Majorana fermions.  In particular, in supersymmetric theories, the lightest supersymmetric particle  is often a Majorana fermion that functions as the dark matter (DM) of the universe  \cite{susy_dm}.  The search for dark matter led to an uptick of interest in the EM interactions of neutral particles, in general, and Majorana fermions, in particular.  Knowledge of these interactions is important for both direct and indirect DM searches and for establishing the relic DM density.  Early constraints on the EM  properties of DM, including the anapole moment and higher-order interactions, can be found in Ref.~\cite{pospelov}.   More recent studies of DM interacting via an anapole moment can be found in Refs.~\cite{Fitzpatrick:2010br,anapole_dm1,anapole_dm2,DelNobile:2014eta}.  Additionally, DM anapole interactions, along with higher order EM interactions, can be studied from the perspective of an effective field theory (EFT); see Ref.~\cite{DeSimone:2016fbz} and the references therein.

In this paper, we will discuss the EM properties of Majorana fermions but move beyond the static limit and focus upon their polarizabilities.
In Sec.~\ref{static}, we first review the well-known static EM properties of the Majorana fermion. In field theory, these static properties  are assessed through interactions mediated by a single low-energy photon.  
Higher-order classical interactions involving induced EM moments, i.e., the particle's polarizability, are quadratic in the EM fields.  As a result, in field theory, these processes involve the interaction between the particle and two photons.  Our first novel results, contained in  Sec.~\ref{2photon}, address these processes.  There, we  determine model-independent constraints on the polarizability of the Majorana fermion that arise as a result of its self-conjugate nature.  Within the context of Compton scattering, we show generally that   two-photon interactions with a Majorana fermion are not forbidden if the process is separately invariant under charge conjugation, parity, and time reversal, but for interactions which are odd under parity, but even under time reversal, the Compton amplitude vanishes.
Finally, in Sec.~\ref{simp_mod}, using a simple model, we explicitly compute the anapole moment of a Majorana fermion along with the leading-order contributions to its polarizabilities.  This section serves as an explicit demonstration of the model-independent results of Sec.~\ref{2photon}; additionally, it suggests a relevant application.  From the explicit model, we show 
that the Majorana fermion is able to undergo $s$-wave annihilation into two photons, mediated by a spin-dependent polarizability, with a cross-section that is larger than a na\"ive EFT estimate might suggest.

\section{Static EM properties \label{static}}

\begin{figure}
\includegraphics[width=4.3cm]{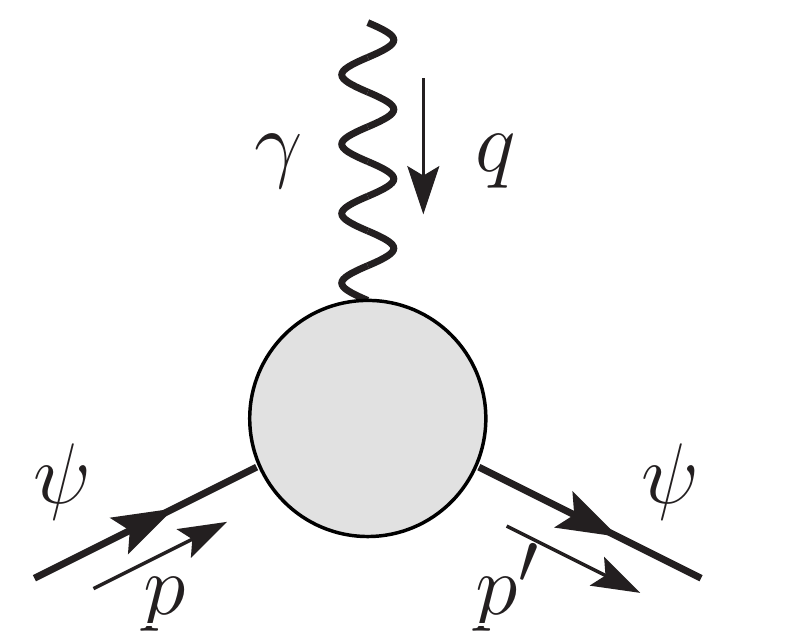}
\caption{ Interaction vertex between an on-shell fermion and an off-shell photon of momentum $q$. \label{fig_vertex}}
\end{figure}

We begin with a review of the known static EM properties of Majorana fermions because the arguments in this section will provide a natural segue to the novel material that follows in Sec.~\ref{2photon}.   To assess the static EM properties of the fermion, we consider the interaction vertex between a single off-shell photon and a spin-$\frac{1}{2}$ fermion  as in Fig.~\ref{fig_vertex}.  
The Lagrangian term mediating this interaction is a contraction of the EM current $J_\text{EM}^\mu = \overline{\Psi} \Gamma^\mu \Psi$ with the photon vector field $A_\mu$.
The structure of this vertex can be ascertained through the  principles of Lorentz covariance and gauge invariance.  
Generally, the matrix element for a conserved EM current operator $J_\text{EM}^\mu$  is characterized by four form factors which are functions of the only nontrivial scalar, the square of the momentum transfer, $q = p'-p$, 
\begin{equation}
\langle \mathbf{p}',s' | J_\text{EM}^\mu | \mathbf{p},s\rangle =  \bar{u}(p')\left[f_1(q^2) \gamma^\mu + \frac{i}{2m} f_2(q^2)  \sigma^{\mu \nu} q_\nu + f_A(q^2) \left( q^2 \gamma^\mu -  q^\mu \qslash \right) \gamma^5 + f_E(q^2) \sigma^{\mu\nu} q_\nu \gamma^5  \right] u(p).  \label{vertex}
\end{equation}
 At zero momentum transfer, these  four functions encode the static EM properties of the fermion.  The first two form factors, $f_1$ and $f_2$, are associated with the charge and anomalous magnetic dipole moment of the fermion.  Both of these terms transform as a classical current four-vector, $J_\text{cl}^\mu$, under parity and time reversal.  Namely, under a parity transformation the spatial part of the current picks up a minus sign $\mathcal{P}J^\mu_\text{cl} (t,\mathbf{x}) \mathcal{P}^{-1} = (-1)^\mu J^\mu_\text{cl} (t,-\mathbf{x}) $ and likewise  under a time reversal $\mathcal{T}J^\mu_\text{cl} (t,\mathbf{x}) \mathcal{T}^{-1} = (-1)^\mu J^\mu_\text{cl} (-t,\mathbf{x})$, where the shorthand $(-1)^\mu$ is defined to be $+1$ for $\mu=0$ and $-1$ for $\mu=j$. 
After the discovery of parity violation in particle physics, Zel'dovich  realized the necessity of the third term in the interaction vertex \cite{zeldovich}.  The anapole moment $f_A(0)$ has mass dimension $[M]^{-2}$, and relative to a classical current, this term is odd under a parity transformation but even under time reversal.  The fourth term is associated with the particle's electric dipole moment.  The function $f_E(q^2)$ has mass dimension $[M]^{-1}$, and this term is odd under both parity and time reversal transformations.  Though a general fermion can have non-zero form factors for all four terms, the EM properties of Majorana particles are highly constrained; only its anapole moment can be nonzero.

Majorana fields are special solutions of the Dirac equation. There is a particular representation of the Dirac matrices that renders all four matrices purely imaginary.  In this Majorana representation,  the solutions to the the Dirac equation are real; the particles represented by these solutions are termed Majorana fermions.  
For other representations of the Dirac matrices, this notion of ``realness" generalizes to the statement that Majorana fields are equal to their Lorentz-covariant conjugates \cite{pal_fermions}.  These fields have half as many degrees of freedom as a spin-$\frac{1}{2}$ Dirac fermion, which translates to the notion that the Majorana fermion is its own antiparticle.
We see that Majorana fermions can be defined without reference to discrete symmetry transformations like charge conjugation, but it is perhaps more intuitive to invoke these ideas in regarding these particles as their own antiparticles.
On the face of it, it might seem that Majorana fermions are eigenstates  of the charge conjugation operator $\mathcal{C}$.  In the absence of any interaction, this is true; however, this notion is of limited utility.  In light of parity violating interactions, the dressed Majorana fermion propagator breaks symmetry under $\mathcal{C}$ \cite{bk82}.
More generally, a Majorana fermion is  defined as a self-conjugate field under  $\mathcal{CPT}$  \cite{bk82,bk83,bk84,pal_fermions}. 
  Because all local Lorentz covariant field theories are invariant under $\mathcal{CPT}$ transformations,  regarding Majorana fermions as their own $\mathcal{CPT}$ conjugates results in no great restriction. Before examining why the EM properties of Majorana fermions are so constrained, we first review how to implement the discrete transformations on spin-$\frac{1}{2}$ fields.

\subsection{Discrete transformations}

Under a parity transformation, the spatial part of a four-vector is reflected; that is, $(t,\mathbf{x}) \mapsto (t, - \mathbf{x})$, or using the shorthand above $x^\mu \mapsto (-1)^\mu x^\mu$.  On the other hand, axial vectors acquire an additional minus sign, relative to vectors, under the parity transformation; as a consequence, a quantity like angular momentum is unchanged under this transformation so that a particle's spin satisfies $\mathbf{s} \mapsto \mathbf{s}$.  This parity transformation of space-time can be implemented by a unitary operator $\mathcal{P}$ on the fermionic field
\begin{equation}
\Psi(t,x) = \int \frac{\mathrm{d}^3p}{(2\pi)^3}\frac{1}{\sqrt{2 E_{\mathbf {p}}}} \sum_s \left( a^s_\mathbf{p} u(\mathbf{p},s) e^{-i p \cdot x}+b^{s \dagger}_\mathbf{p} v(\mathbf{p},s) e^{i p \cdot x} \right) ,  \label{field}
\end{equation}  
where $a^s_\mathbf{p}$ and $b^s_\mathbf{p}$ are annihilation operators for fermions and anti-fermions, respectively. Acting on the field operators, a parity transform reverses the momentum direction for the annihilation operators; e.g., we have $\mathcal{P} a^s_\mathbf{p} \mathcal{P}^{-1} = a^s_\mathbf{-p}$, up to a phase which does not impact our considerations in this work. 
Employing the Weyl representation of the Dirac matrices, spinors have the property
\begin{eqnarray}
u(\mathbf{p},s) &=& \gamma^0 u(-\mathbf{p},s),\label{p_uspinor}\\
v(\mathbf{p},s) &=& - \gamma^0 v(-\mathbf{p},s), \label{p_vspinor}
\end{eqnarray}
so that the parity transformation on the fields can be taken as $\mathcal{P} \Psi(t,\mathbf{x}) \mathcal{P}^{-1} =i \gamma^0 \Psi(t,-\mathbf{x})$. 

Under charge conjugation, particles are mapped to antiparticles while maintaining the same momentum and spin.  That is, the action of the charge conjugation operator maps fermion annihilation operators to anti-fermion annihilation operators, $\mathcal{C} a^s_\mathbf{p} \mathcal{C}^{-1} = b^s_\mathbf{p}$, up to a phase.  
To determine the impact of this map on the field $\Psi(t,\mathbf{x})$, Eq.~(\ref{field}), we introduce the unitary charge conjugation matrix $C$.  The defining feature of this matrix is its action upon the Dirac matrices, namely $C \gamma_\mu C^{-1} = - \gamma_\mu^\top$ where $\top$ denotes the transpose.  In addition to its unitarity, the charge  conjugation matrix is antisymmetric, $C^\top = -C$.  In the Weyl representation of the Dirac matrices, the charge conjugation matrix can be taken as $C = -i \gamma^2 \gamma^0$, and it provides a map between $u$ and $v$ spinors
\begin{eqnarray}
u(\mathbf{p},s) &=& C\gamma^0 v^*(\mathbf{p},s),  \label{c_uspinor}\\
v(\mathbf{p},s) &=& C\gamma^0 u^*(\mathbf{p},s). \label{c_vspinor}
\end{eqnarray}
 With this map, the (linear) charge conjugation operator satisfies $\mathcal{C} \Psi(t,\mathbf{x})\mathcal{C}^{-1} =C \gamma^0 {\Psi}^*(t,\mathbf{x}) = C \overline{\Psi}^\top(t,\mathbf{x})$. 

In what follows, we will be particularly interested in the transformation of Dirac bilinears under charge conjugation.  Extending the work of the previous paragraph, we find the effect of $\mathcal{C}$ on the Dirac adjoint of the fermion field to be $\mathcal{C} \overline{\Psi}(t,\mathbf{x}) \mathcal{C}^{-1} = \Psi(t,\mathbf{x})^\top C$.  Given this, a Dirac bilinear will transform as $\mathcal{C} \overline{\Psi} A \Psi \mathcal{C}^{-1} =\Psi(t,\mathbf{x})^\top C A C \overline{\Psi}^\top$, where $A$ is some element of the Dirac algebra.  We can put this expression in a more useful form by taking its transpose; recalling that we acquire a minus sign upon the anticommutation of the fermion fields, we find $\mathcal{C} \overline{\Psi} A\Psi \mathcal{C}^{-1}  = \overline{\Psi} C A^\top C^{-1} \Psi$. Given the defining property of the charge conjugation matrix, $C (\gamma^\mu)^\top C^{-1} = - \gamma^\mu$, we can establish the following relationships
\begin{eqnarray}
C[ \gamma^5 ]^\top  C^{-1}& =& \gamma^5, \label{Cg5C}\\
C[ \sigma^{\mu\nu} ]^\top  C^{-1}& =& -\sigma^{\mu\nu},\\
C[ \gamma^\mu \gamma^5 ]^\top  C^{-1}& =& \gamma^\mu \gamma^5,\\
C[ \sigma^{\mu\nu} \gamma^5 ]^\top  C^{-1}& =& -\sigma^{\mu\nu} \gamma^5.
\end{eqnarray}  
We see, for instance, that the first term in the EM current, Eq.~(\ref{vertex}), is $\mathcal{C}$-odd, $\mathcal{C} \overline{\Psi} \gamma^\mu \Psi \mathcal{C}^{-1} = - \overline{\Psi}\gamma^\mu \Psi$.  In addition to elements of the Dirac algebra, we also find momentum-dependent terms in Eq.~(\ref{vertex}) which signify the presence of a derivative coupling in the effective Lagrangian for the EM vertex.  Charge conjugation commutes with derivatives, but when transposing fields in a bilinear, the derivative will shift positions; e.g., $\mathcal{C} \overline{\Psi} A(\partial_\nu \Psi)\mathcal{C}^{-1} = (\partial_\nu \overline{\Psi}) C \Gamma^\top C^{-1} \Psi$.  As an example, we can confirm that the anomalous magnetic moment term in Eq.~(\ref{vertex}) is $\mathcal{C}$-odd
\begin{equation}
\mathcal{C}[ (\partial_\nu \overline{\Psi}) \sigma^{\mu\nu} \Psi + \overline{\Psi} \sigma^{\mu\nu} (\partial_\nu \Psi) ]\mathcal{C}^{-1} =- \overline{\Psi} \sigma^{\mu\nu} (\partial_\nu \Psi) - (\partial_\nu \overline{\Psi}) \sigma^{\mu\nu} \Psi.
\end{equation}
Pursuing this example, we compute the matrix element, modulo some factors, for this operator assuming initial and final fermion states with momenta $\mathbf{p}$ and $\mathbf{p}'$ and spins $s$ and $s'$, respectively
\begin{equation}
\langle 0 | a^{s'}_{\mathbf{p}'} [(\partial_\nu \overline{\Psi}) \sigma^{\mu\nu} \Psi + \overline{\Psi} \sigma^{\mu\nu} (\partial_\nu \Psi)] a^{s\dagger}_\mathbf{p} |0 \rangle \sim i (p_\nu' - p_\nu) \bar{u}(\mathbf{p}',s') \sigma^{\mu \nu} {u}(\mathbf{p},s).
\end{equation}
For completeness, we execute the same computation for anti-fermions
\begin{equation}
\langle 0 | b^{s'}_{\mathbf{p}'} [(\partial_\nu \overline{\Psi}) \sigma^{\mu\nu} \Psi + \overline{\Psi} \sigma^{\mu\nu} (\partial_\nu \Psi)] b^{s\dagger}_\mathbf{p} |0 \rangle \sim -i (-p_\nu + p_\nu'  ) \bar{v}(\mathbf{p},s) \sigma^{\mu \nu} {v}(\mathbf{p}',s').
\end{equation}
To determine the matrix element for anti-fermions, we see that an  overall minus sign arises from the anticommutation of fermion fields, and  momentum-dependent factors can be got from the fermion amplitude via the substitution $p \mapsto -p'$ and $p' \mapsto -p$.

The final discrete transformation that we discuss is time reversal.  For this transformation, the temporal part of a spacetime vector acquires a minus sign, $(t,\mathbf{x}) \mapsto (-t,  \mathbf{x})$, and the direction of angular momentum is flipped $\mathbf{s} \mapsto -\mathbf{s}$.  To implement this  on a fermionic field, the transformation must be conjugate linear; that is, for a scalar $\alpha$, the transformation conjugates the scalar $\mathcal{T}\alpha \Psi \mathcal{T}^{-1} = \alpha^* \mathcal{T} \Psi \mathcal{T}^{-1}$.   In the Weyl representation, we note that spinors of opposite spin can be related via
\begin{eqnarray}
u(-\mathbf{p},-s) &=& -\gamma^1 \gamma^3 u^*(\mathbf{p},s),  \label{t_uspinor}\\
v(-\mathbf{p},-s) &=& -\gamma^1 \gamma^3 v^*(\mathbf{p},s).  \label{t_vspinor}
\end{eqnarray}
So, we can implement the time reversal operator  on fields as  $\mathcal{T}  \Psi(t,\mathbf{x}) \mathcal{T}^{-1} = (-\gamma^1 \gamma^3) \Psi(-t,\mathbf{x}) =- C \gamma^5 \Psi(-t,\mathbf{x})$.

We can combine all these transformations to determine the joint effect of the anti-unitary operator $\mathcal{CPT}$; we find $(\mathcal{CPT}) \Psi(t,\mathbf{x}) (\mathcal{CPT})^{-1} = -i \gamma^5 \gamma^0 \overline \Psi^\top(-t,-\mathbf{x})$.  Combining the three relationships amongst the spinors in Eqs.~(\ref{p_uspinor}--\ref{c_vspinor},\ref{t_uspinor}--\ref{t_vspinor}), we find the useful relationship
\begin{eqnarray}
u(\mathbf{p},-s) &=& -\gamma^5 v(\mathbf{p},s),\\
v(\mathbf{p},-s) &=&\gamma^5 u(\mathbf{p},s).
\end{eqnarray}
Of interest is how Dirac bilinears transform under $\mathcal{CPT}$.  To evaluate these bilinears, we must determine the action of $\mathcal{CPT}$ on the Dirac adjoint of the field, $(\mathcal{CPT}) \overline{\Psi}(t,\mathbf{x}) (\mathcal{CPT})^{-1} = -i \Psi(-t,-\mathbf{x})^\top   \gamma^0 \gamma^5$.  Then, for the operator $A$, we find that the bilinear transforms is
\begin{equation}
(\mathcal{CPT}) \overline{\Psi} A \Psi (\mathcal{CPT})^{-1} = - \Psi^\top \gamma^0 \gamma^5 A^*  \gamma^5 \gamma^0 \overline{\Psi}^\top.
\end{equation}
Taking the transpose of the above number and introducing a minus sign upon anticommutation of the fermion fields, this simplifies to 
\begin{equation}
(\mathcal{CPT}) \overline{\Psi} A \Psi (\mathcal{CPT})^{-1} =  \overline{\Psi} \gamma^5  A  \gamma^5  \Psi  \label{cpt}
\end{equation}
where we have assumed Hermiticity, $\gamma^0 A^\dagger \gamma^0 = A$.  Using Eq.~(\ref{cpt}), we are able to determine that the EM fermionic current $J^\mu_\text{EM}$ is odd under $\mathcal{CPT}$.  This could be ascertained in another manner.  Noting that the photon field $A_\mu$ is odd under $\mathcal{CPT}$, then the EM current must also be odd in order for this local QFT to be invariant under $\mathcal{CPT}$.

\subsection{Constraints on the amplitude}

Knowing how the current transforms under $\mathcal{CPT}$ will help us constrain the EM form factors of a Majorana fermion.  To do so,  we recall that Majorana fermions are  their own $\mathcal{CPT}$ conjugates; that is, a state transforms as $\mathcal{CPT} | \mathbf{p},s\rangle = | \mathbf{p}, -s\rangle$.   Under an anti-unitary transformation $\widetilde {U}$, the entries in an inner product are switched, i.e., $\langle \widetilde U a | \widetilde U b \rangle =\langle b | a \rangle  $.  Because the $\mathcal{CPT}$ transformation is anti-unitary, then the matrix element for the EM current  for a Majorana fermion satisfies
\begin{equation}
\langle \mathbf{p}',s' | J_\text{EM}^\mu | \mathbf{p},s\rangle = - \langle \mathbf{p}, -s | J_\text{EM}^\mu | \mathbf{p}',-s'\rangle,  \label{cpt_jem}
\end{equation}
where the minus sign arises for the transformation properties of $J_\text{EM}^\mu$ under $\mathcal{CPT}$.  The left-hand side (LHS) of this equation can be expanded as above in Eq.~(\ref{vertex}).  The right-hand side (RHS) of Eq.~(\ref{cpt_jem}) follows suit
\begin{equation}
\langle \mathbf{p}, -s | J_\text{EM}^\mu | \mathbf{p}',-s'\rangle = \bar{u}(\mathbf{p},-s) \Gamma^\mu(p,p') u(\mathbf{p}',-s'),
\end{equation}
but we can use the relations in Eqs.~(\ref{p_uspinor}) and (\ref{t_uspinor}) to find an alternate expression for the spinors $u(\mathbf{p},-s) = -\gamma^0 C \gamma^5 u^*(\mathbf{p},s)$.  Using this expression and the related one for the Dirac adjoint, we can rewrite the RHS of Eq.~(\ref{cpt_jem}) as
\begin{eqnarray}
 \bar{u}(\mathbf{p},-s) \Gamma^\mu(p,p') u(\mathbf{p}',-s')&=&- u^\top(\mathbf{p},s) \gamma^5 C \Gamma^\mu(p,p') \gamma^0 C \gamma^5 u^*(\mathbf{p}',s')  \label{bit}\\
&=& \bar{u}(\mathbf{p}',s') \gamma^5 C [\Gamma^\mu(p,p')]^\top  C^{-1} \gamma^5 u(\mathbf{p},s), \label{final}
\end{eqnarray}
where we take the transpose of the number on the RHS of Eq.~(\ref{bit}) to arrive at the final expression.  
Noting  $p-p' := -q$, then we find 
\begin{equation}
\gamma^5 C [\Gamma^\mu(p,p')]^\top  C^{-1} \gamma^5 = f_1(q^2) \gamma^\mu + \frac{i}{2m} f_2(q^2)  \sigma^{\mu \nu} q_\nu - f_A(q^2) \left( q^2 \gamma^\mu -  q^\mu \qslash \right) \gamma^5 +  f_E(q^2) \sigma^{\mu\nu} q_\nu \gamma^5. \label{gampp'}
\end{equation}
Returning to Eq.~(\ref{cpt_jem}) where we first implemented the $\mathcal{CPT}$ transformation,  we find in light of Eq.~(\ref{gampp'}) that only the Majorana fermion's anapole is non-vanishing; that is, $f_1, f_2, f_E \equiv 0$.

These restrictions upon the static EM properties of the Majorana fermion can be understood heuristically by considering the non-relativistic limit of the field theory.  This argument originally appeared in Refs.~\cite{bk82,nieves} and bears repeating given its insightful simplicity.  Clearly, a Majorana fermion must carry no electric charge if it is to be its own antiparticle, but why must both its magnetic and electric dipole moments vanish?  In the low energy limit, the interaction Hamiltonian between the EM field and a particle's magnetic, $\mu$, and electric, $d$, dipole moments is $H_\text{int}  = -  \mu\, ({\mathbf{s}} \cdot {\mathbf{B}}) -  d\, ( {\mathbf{s}} \cdot {\mathbf{E}} )$ where we denote the particle's spin by $ {\mathbf{s}}$.  Spin is odd under $\mathcal{CPT}$ whereas the electric and magnetic fields are both even under the transformation.  If our the interaction is $\mathcal{CPT}$ invariant, then the magnetic and electric dipole must vanish.  On the other hand, the non-relativistic Hamiltonian for the term involving the anapole moment $a$ is $H_\text{int}  = -  a\, ({\mathbf{s}} \cdot {\mathbf{J}})$, because the EM four-current is given by the divergence of the EM tensor $\partial_\mu F^{\mu \nu} = J^\nu$.  The current ${\mathbf{J}}$ is odd under $\mathcal{CPT}$ so that overall this interaction is invariant under $\mathcal{CPT}$.  

\section{Two photon interactions \label{2photon}}

We turn our attention to processes involving fermions and two real photons.  As an exemplar, we consider the extensively studied process of real Compton scattering.  The low energy limit of the Compton scattering process is determined by the Born contribution computed via  tree-level Feynman diagrams involving only a single virtual fermion with real photons coupling via the above vertex, Eq.~(\ref{vertex}) and Fig.~\ref{fig_vertex}.  As such, in a low-energy expansion, the leading order contribution to the Compton scattering amplitude is determined exclusively by the static EM properties of the fermion \cite{low,lapidus,brodsky,gg}.  But, as shown above, the only non-zero static EM property of a Majorana fermion is its anapole moment, and the coupling between this anapole moment and real photons vanishes.  Thus, for Majorana fermions, the typical leading order contributions to the Compton amplitude will vanish.  

But, moving beyond the Born approximation, there are model-dependent corrections to the amplitude, namely electric and magnetic polarizabilities, that are relevant at higher energies. Majorana fermions are not forbidden from interacting with photons through such induced electric and magnetic dipole moments \cite{radescu}.  Two-photon processes are the simplest avenue to explore induced moments; heuristically, one photon can be thought of as inducing an EM moment which can then interact with the other photon.  These higher-order corrections to the Compton amplitude have been well studied, particularly for nucleons \cite{prange, jacob_mathews, hearn, bardeen, Bernard:1991rq, hhkk, babusci, holstein_hopol, bedaque, chen, Bernard:2002pw, Hildebrandt:2003md, holstein_sumrules, gorchtein},  and though the precise details of a particle's polarizabilities are model dependent, the general framework for Compton scattering established in these previous studies is relevant for Majorana fermions.  We begin our exploration of two-photon processes by constructing a manifestly covariant expression for the  Compton scattering amplitude for a general spin-$\frac{1}{2}$ particle, including both the $\mathcal{P}$-even,  $\mathcal{T}$-even  and $\mathcal{P}$-odd,  $\mathcal{T}$-even contributions.  Supposing a Majorana fermion is the scatterer, we then determine what additional constraints this imposes  upon the structure of the  amplitude. In the end, we show that the existence of electric and magnetic polarizabilities is not forbidden for Majorana fermions.  After the formal manipulations are  complete, we conclude this section with a more intuitive, non-relativistic discussion of the polarizabilities.

\subsection{Manifestly covariant amplitude }

\begin{figure}
\includegraphics[width=4.3cm]{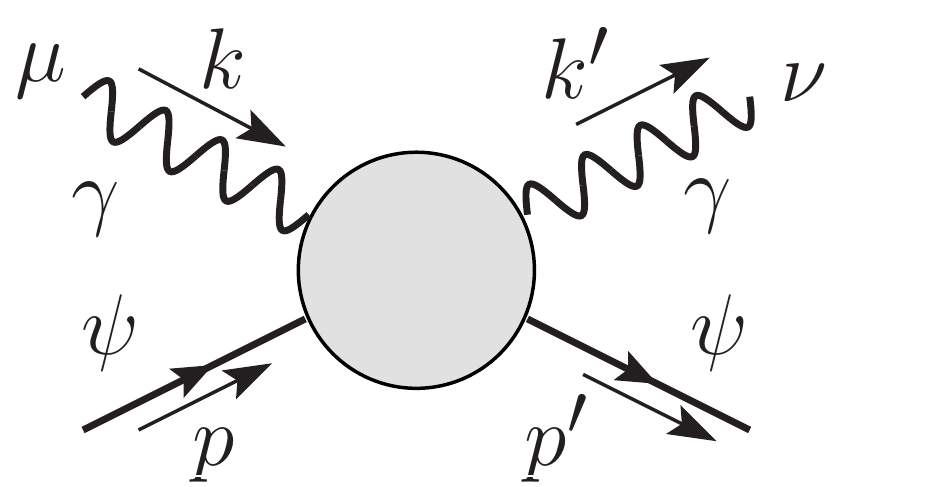}
\caption{ Two photon interaction with a fermion.  The photons are assumed to be real and transverse, and the fermion is on mass shell. \label{fig_compton}}
\end{figure}

In Fig.~\ref{fig_compton}, we lay out the kinematics for a real photon scattered by a general spin-$\frac{1}{2}$ fermion.   The incoming photon momentum, $k$, is on shell $k^2=0$, and the photon polarization, $\epsilon$, is transverse, $\epsilon \cdot k =0$.  Similar relations hold for the outgoing photon's momentum $k'$ and polarization vector $\epsilon'$.  The incoming and outgoing fermions are on mass shell, $p^2 = p'^2 = m^2$.  The structure of the amplitude for real Compton scattering is $\mathcal{M} = {\epsilon_\nu'}^* \epsilon_\mu \bar{u}(\mathbf{p}',s') \Gamma^{\nu \mu}(p',k';p,k) u(\mathbf{p},s)$.  Following Ref.~\cite{prange}, we will construct a manifestly covariant expression for the tensor $\Gamma^{\nu\mu}$ which characterizes the Compton amplitude.  Using crossing symmetry and assuming separate invariance under the discrete symmetries $\mathcal{C}$, $\mathcal{P}$, and $\mathcal{T}$, the tensor  can be decomposed into six terms \cite{prange}.  
Because parity violation is a necessary ingredient for the existence of anapole moment, we extend the work of Ref.~\cite{prange}  to show that the parity-violating (but $\mathcal{T}$-even) portion of the Compton scattering tensor can be decomposed into four terms.

We construct a covariant representation of the tensor $\Gamma^{\nu\mu}$ from the set of orthogonal basis vectors:
$ K^\mu = \frac{1}{2}(k^\mu+k'^\mu)$, $P'^\mu = P^\mu-\frac{P\cdot K}{K^2}K^\mu$, $q^\mu = p'^\mu - p^\mu $, and $N^\mu = \epsilon^{\mu\nu\rho\sigma} P'_\nu K_\rho q_\sigma$  where $P^\mu = \frac{1}{2}(p'^\mu+p^\mu)$ \cite{prange}.
By exploiting Lorentz covariance, transverse polarization, and the Ward identity, $\Gamma^{\nu\mu} k_\mu=0=\Gamma^{\nu\mu} k'_\nu$, we can constrain the structure of the tensor so that it is a linear combination of the following four tensors: $P'^\mu P'^\nu$, $N^\mu N^\nu$, $P'^\mu N^\nu\pm N ^\mu P'^\nu$.  The coefficients of these tensors each contain, as a factor, a bilinear $\bar{u}(\mathbf{p}',s') A u(\mathbf{p},s)$ where $A$ is $\mathbbm{1}, \Kslash,\gamma^5$, or $\Kslash \gamma^5$.  If the amplitude is to be invariant under parity, then the tensors $P'^\mu P'^\nu$ and $N^\mu N^\nu$ will carry the scalar coefficients formed whenever the operator $A$ is $\mathbbm{1}$ or 
$\Kslash$, whereas the remaining tensors $P'^\mu N^\nu\pm N ^\mu P'^\nu$ require pseudoscalar coefficients with $A=\gamma^5$ or $\Kslash \gamma^5$.  This results in eight different terms. 

These eight terms can be further reduced to six terms by requiring that the amplitude  be invariant under charge conjugation and by exploiting crossing symmetry.  Again, the matrix elements for the two-photon operator involves factors of momenta, $p^\mu$ and $p'^\mu$, signaling derivative couplings between the fermion fields and photons.  As outlined in the previous section, we effect the charge conjugation via
\begin{equation}
\mathcal{C}   \Gamma^{\nu \mu}(p',k';p,k)  \mathcal{C}^{-1} = C [\Gamma^{\nu \mu}(-p,k';-p',k)]^\top C^{-1}. \label{c_cond}
\end{equation}
As noted in Ref.~\cite{prange}, under the action of  charge conjugation alone, it is difficult to arrive at additional constraints on the structure of the Compton amplitude because the Mandelstam variables change under this map, $s \mapsto u$ and $u \mapsto s$.  But, if one makes use of crossing symmetry for the photons, namely $k\leftrightarrow -k'$ and $\mu \leftrightarrow \nu$, in conjunction with charge conjugation, then it is easy to see that only six of the eight terms in the decomposition of $\Gamma^{\nu\mu}$ are $\mathcal{C}$-even. Implementing crossing symmetry, the condition for being even under $\mathcal{C}$ can be expressed as
\begin{equation}
\Gamma^{\nu \mu}(p',k';p,k) = C [\Gamma^{\mu \nu}(-p,-k;-p',-k')]^\top C^{-1}. 
\end{equation}
Under this joint map, the basis vectors $P'$ and $K$ are odd while $q$ and $N$ are even.
In the end, assuming separate invariance under $\mathcal{C}$, $\mathcal{P}$, and $\mathcal{T}$, the tensor can be decomposed as
\begin{equation}
\Gamma^{\nu\mu} =(T_1 \mathbbm{1} + T_4  \Kslash)\frac{P'^\mu P'^\nu}{P'^2}+ (T_2\mathbbm{1} + T_5  \Kslash)\frac{N^\mu N^\nu}{N^2} + iT_3 \gamma^5 \frac{(P'^\mu N^\nu- N ^\mu P'^\nu)}{\sqrt{P'^2N^2}}  + iT_6  \gamma^5 \Kslash \frac{(P'^\mu N^\nu+ N ^\mu P'^\nu)}{\sqrt{P'^2N^2}} ,  \label{gam_munu}
\end{equation}
where the $T_j$ are functions of the Mandelstam invariants and we omit the spinors $\bar{u}(\mathbf{p}',s')$ and $u(\mathbf{p},s)$ to condense notation \cite{prange}. (Note, we have opted to construct our tensors using normalized vectors as in Ref.~\cite{hearn}.)

Moving beyond the work in Ref.~\cite{prange}, we also consider  contributions to the Compton amplitude which are $\mathcal{P}$-odd and $\mathcal{T}$-even (and thus $\mathcal{C}$-odd). The tensor structure for $\Gamma\Podddown^{\nu\mu}$ derives from the same set of four tensors; however, in order to have $\mathcal{P}$-odd terms, the $P'^\mu P'^\nu$ and $N^\mu N^\nu$ tensors  require pseudoscalar coefficients, and $P'^\mu N^\nu\pm N ^\mu P'^\nu$ require scalar coefficients. 
If we use crossing symmetry and require the amplitude to be $\mathcal{C}$-odd, then we arrive at the following condition
\begin{equation}
\Gamma\Podddown^{\nu \mu}(p',k';p,k) = -C [\Gamma\Podddown^{\mu \nu}(-p,-k;-p',-k')]^\top C^{-1}.
\end{equation}
We find four terms in the amplitude which are $\mathcal{P}$-odd and $\mathcal{T}$-even
\begin{equation}
\Gamma\Podddown^{\nu\mu} = T\Podddown_1 \gamma^5\Kslash  \frac{P'^\mu P'^\nu}{P'^2}  + T\Podddown_2 \gamma^5\Kslash  \frac{N^\mu N^\nu}{N^2}  + (T\Podddown_3 \mathbbm{1}+ T\Podddown_4  \Kslash) \frac{(P'^\mu N^\nu+ N ^\mu P'^\nu)}{\sqrt{P'^2N^2}} .\label{gamPx_munu}
\end{equation}
As an aside, we note that one can refine the expressions for the covariant tensors, freeing them of both kinematic singularities and zeros \cite{bardeen}, but this is not necessary for our purposes.

\subsection{Center-of-momentum frame}

Before we discuss the restrictions that a Majorana scatterer will place upon the tensor decomposition, Eqs.~(\ref{gam_munu}) and (\ref{gamPx_munu}), we will first connect this manifestly covariant expression with a more familiar frame-specific representation of the scattering amplitude.  We consider the center-of-momentum (CoM) frame where the photon 4-momenta are given by $k=(\omega, \omega \khat)$ and $k' = (\omega, \omega \khat')$ with a scattering angle $\theta$ that satisfies $\khat \cdot \khat' = \cos \theta$.  For completeness, the fermion 4-momenta are $p = (E, - \omega \khat)$ and $p'=(E, -\omega \khat')$ with $E=\sqrt{m^2+\omega^2}$ for fermion mass $m$.  In this frame, the Compton amplitude can be simply expressed in terms of Pauli spinors  for the initial (final) fermion state $\xi$ ($\xi'$).  Omitting these spinors, the portion of the amplitude even under $\mathcal{C}$, $\mathcal{P}$, and $\mathcal{T}$, i.e., Eq.~(\ref{gam_munu}), can be written as 
\al{
\mathcal{M} =& A_1\, (\beps'^* \cdot \beps) + A_2\, (\beps'^*\cdot \khat)  (\beps \cdot \khat') + i A_3\, [\bsig \cdot (\beps \times \beps'^* )] +i A_4\, [\bsig \cdot (\khat \times \khat')] (\beps'^* \cdot \beps) \nonumber \\&
+ i A_5 \, \bsig \cdot [(\beps'^*\times \khat) (\beps \cdot \khat') - (\beps \times \khat') (\beps'^* \cdot \khat)] + i A_6\, \bsig \cdot [(\beps'^* \times \khat') (\beps \cdot \khat')- (\beps \times \khat)(\beps'^*\cdot \khat) ] \label{amp_com}
}
where the coefficients are functions of the photon energy and scattering angle $A_j(\omega,\theta)$ \cite{jacob_mathews}.
A map between the covariant expression, Eq.~(\ref{gam_munu}), and the CoM amplitude, Eq.~(\ref{amp_com}), was constructed in Ref.~\cite{hearn}
\al{
A_1 = & c_1 T_2  -c_2 T_5 \label{A1},\\
A_2 = & \frac{1}{\sin^2\theta} [c_1 (T_1 + \cos\theta T_2) -c_2 (T_4 + \cos \theta T_5) ],  \\
A_3 = & (E-m) [T_1+\cos\theta T_2 -(E+m+\omega)(T_4 +\cos\theta T_5)]  +2\omega [T_3 - (E+\omega)T_6],\\
A_4 = & (E-m)[T_2 -(E+m+\omega)T_5],\\
A_5 = &  -\frac{E-m}{\sin^2\theta} \cos \theta [(T_1+ \cos \theta T_2) -(E+m+\omega)(T_4+\cos \theta T_5)  ]- \frac{\omega}{\sin^2\theta} [(1+\cos \theta)T_3  +(E+\omega) (1-\cos \theta)T_6],\\
A_6 =&\frac{E-m}{\sin^2\theta}[ T_1 +\cos \theta T_2 -(E+m+\omega)(T_4 +\cos \theta T_5)]  + \frac{\omega}{\sin^2\theta}[(1+\cos \theta)T_3 -(E+\omega) (1-\cos \theta)T_6]. \label{A6}
}
We define $c_1 = E+m-(E-m) \cos \theta$ and $c_2 = (E+m)(m-E-\omega)-(E-m)(m+E+\omega) \cos \theta$.
[NB:  Our expressions for $A_j$ in terms of $T_j$ differ slightly from those in Ref.~\cite{hearn} due to differing choices in conventions.]

In Ref.~\cite{gorchtein}, we find the forward Compton amplitude in the CoM frame 
\al{\mathcal{M}\Podddown =& A\Podddown_1\, (\beps'^*\cdot \beps) [\bsig\cdot(\khat'+\khat) ]+ A\Podddown_2\, (\beps'^*\cdot \khat)(\beps\cdot \khat') [\bsig\cdot(\khat'+\khat) ] \nonumber \\
&+ iA\Podddown_3\, (\khat'+\khat)\cdot[\beps'^*\times\beps] + iA\Podddown_4\, \bsig\cdot[(\khat-\khat')\times(\beps'^*\times \beps) ], \label{amp_podd_com}
}
where $A\Podddown_j = A\Podddown_j(\omega,\theta)$ and we omit the Pauli spinors.  We can construct a map from our covariant expression for the $\mathcal{P}$-odd, $\mathcal{T}$-even portion of the Compton amplitude, Eq.~(\ref{gamPx_munu}), to the CoM frame via
\al{
A\Podddown_1 = &  \omega(E+\omega)   T\Podddown_2 \label{A1Px},\\
A\Podddown_2 = &  \frac{1}{\sin^2 \theta} \{ \omega(E+\omega)[T\Podddown_1 +  \cos \theta  T\Podddown_2 ]
  +i(1-\cos \theta)(E -m ) [    T\Podddown_3 - (E+m+\omega)  T\Podddown_4 ]\},\\
A\Podddown_3 = & -i \frac{(1-\cos \theta)}{\sin^2\theta} [c_1 T\Podddown_3 +c_2   T\Podddown_4],\\
A\Podddown_4 =&   (E -m )  [ T\Podddown_3 - (E+m+\omega) T\Podddown_4]  \label{A4Px}.
}

With these connections between the covariant and CoM expressions, we can more easily make contact with the leading order contributions to the polarizabilities of the Majorana fermion by executing a low energy expansion in the CoM frame \cite{ragusa, gorchtein}.

\subsection{Constraints on the amplitude }

Now, we address any constraints that the Majorana character of the scatterer might have upon the structure of the Compton amplitude tensor.  In this case, it is more useful to exploit the self-conjugate nature of the Majorana fermion under Lorentz-covariant conjugation, rather than appeal to self-conjugacy under $\mathcal{CPT}$.  A fermionic field is Majorana if $\Psi = \gamma^0 C \Psi^*$
 \cite{pal_fermions}; as a consequence, there is  nothing to distinguish particle from antiparticle in the field expansion
 \begin{equation}
\Psi(t,x) = \int \frac{\mathrm{d}^3p}{(2\pi)^3}\frac{1}{\sqrt{2 E_{\mathbf {p}}}} \sum_s \left( a^s_\mathbf{p} u(\mathbf{p},s) e^{-i p \cdot x}+a^{s \dagger}_\mathbf{p} v(\mathbf{p},s) e^{i p \cdot x} \right).  \label{maj_field}
\end{equation}  
Given this, the Compton scattering amplitude for Majorana fermions is
\begin{equation}
\mathcal{M} = {\epsilon_\nu'}^* \epsilon_\mu\left\{ \bar{u}(\mathbf{p}',s') \Gamma^{\nu \mu}(p',k';p,k) u(\mathbf{p},s)
-\bar{v}(\mathbf{p},s) \Gamma^{\nu \mu}(-p,k';-p',k) v(\mathbf{p}',s') \right\}.
\end{equation}
But, using Eq.~(\ref{c_vspinor}), we can rewrite this as
\begin{equation}
\mathcal{M} = {\epsilon_\nu'}^* \epsilon_\mu \bar{u}(\mathbf{p}',s') [\Gamma^{\nu \mu}(p',k';p,k) + C\Gamma^{\nu \mu}(-p,k';-p',k)^\top C^{-1} ]u(\mathbf{p},s).
\end{equation}
We recall the effective action of the charge conjugation operator on $\Gamma^{\nu\mu}$, Eq.~(\ref{c_cond}).  From this, we find that if the amplitude is $\mathcal{C}$-even then $\mathcal{M} = 2 {\epsilon_\nu'}^* \epsilon_\mu  \bar{u}(\mathbf{p}',s') \Gamma^{\nu \mu}(p',k';p,k) u(\mathbf{p},s)$; that is,  the amplitude for a Majorana fermion is double the expected value for a Dirac fermion.  But, if the amplitude is $\mathcal{C}$-odd, then it vanishes identically.  The fact that the Majorana fermion is self-conjugate means that particle cannot be distinguished from antiparticle, forcing any $\mathcal{C}$-odd portions to vanish.

By examining the Compton amplitude at low energies, we will be able to understand the existence of the Majorana polarizabilities from a classical standpoint.  
We focus on the electric polarizability of a substance consisting of  spherically symmetric atoms.  The atom has no permanent electric dipole moment, but when placed in an external electric field, opposite forces on the protons and electrons will create  a nonzero dipole moment parallel to the external field.  At leading order, this dipole moment is proportional to the field strength,  $\mathbf{d} = 4 \pi \alpha \mathbf{E}$, where $\alpha$ is the spin-independent electric polarizability.  In terms of energetics, the dielectric medium contributes in the usual way, $H_\text{int} = - 2 \pi \alpha E^2$.   A similar construction applies for the magnetic field with an induced dipole moment, $\boldsymbol \mu = 4 \pi \beta \mathbf{B}$, characterized by the spin-independent magnetic polarizability $\beta$.   Together, the classical interaction Hamiltonian for the spin-independent polarizabilities is $H_\text{int} = -2\pi \alpha E^2 - 2\pi \beta B^2$.  

At low energies, the field theoretic scattering amplitude should mirror the classical interactions.  The Hamiltonian is quadratic in the fields, and we can connect this with the field theoretic amplitude by identifying one occurrence of the field with the incident photon and the other with the outgoing photon.  Writing the interaction Hamiltonian in terms of the electromagnetic tensor, this becomes $H_\text{int} \sim  \alpha  F^{0j}F'_{0j} -  \tfrac{1}{2} \beta F^{jk}F'_{jk}$.  We assume a  classical plane-wave vector potential, $\mathbf{A} = A_0 \beps\, e^{i\omega (\khat \cdot \mathbf{x} -t)}$, for the $k$th mode of the electric and magnetic fields.
Then focusing on, say, the magnetic polarizability term in the CoM frame, we find $H_\text{int}  \sim -\beta \omega^2[(\khat' \cdot \khat) (\beps'^* \cdot \beps) - (\beps'^* \cdot\khat)(\beps \cdot \khat') ]$.  At low energies, we expect the magnetic spin-independent polarizability to appear in the Compton amplitude, Eq.~(\ref{amp_com}), via $A_1 \sim\cos\theta \beta  \omega^2 $ and $A_2 \sim - \beta \omega^2$.
In fact, this is the case.  A low energy expansion (LEX) of the Compton amplitude in the CoM frame for the $\mathcal{P}$-even, $\mathcal{T}$-even contribution to the scattering amplitude can be found in Ref.~\cite{hhkk}.  We reproduce that expansion here omitting terms that are $\mathcal{O}(\omega^4)$
\al{
A_1(\omega, \theta) &\approx 8 \pi m_\chi (\alpha + \cos \theta \, \beta) \omega^2+8 \pi (\alpha + \beta) (1+ \cos \theta)\omega^3  \label{A1_LEX} \\
A_2(\omega, \theta) &\approx -8 \pi m_\chi  \beta \omega^2-8 \pi (\alpha + \beta) \omega^3  \label{A2_LEX}  \\
A_3(\omega, \theta) &\approx -8\pi m_\chi [ \gamma_1 -(\gamma_2 + 2 \gamma_4) \cos \theta] \omega^3  \label{A3_LEX}   \\
A_4(\omega, \theta) &\approx -8\pi m_\chi \gamma_2 \omega^3 \\
A_5(\omega, \theta) &\approx 8 \pi m_\chi \gamma_4 \omega^3 \\
A_6(\omega, \theta) &\approx 8\pi m_\chi \gamma_3 \omega^3. \label{A6_LEX}
 }  
In addition to $\alpha$ and $\beta$, we find the appearance of four spin-dependent polarizabilities, $\gamma_j$ \cite{ragusa}.  There are two electric polarizabilities $\mathbf{d} = 4\pi \gamma_1[\mathbf{s} \times (\boldsymbol{\nabla}\times \mathbf{B})]$  and $\mathbf{d} = 4\pi \gamma_3 \boldsymbol{\nabla}(\mathbf{s}\cdot \mathbf{B})$ induced by a non-uniform magnetic field, and two magnetic polarizabilities $\mathbf{m} = 4\pi \gamma_2 \boldsymbol{\nabla}(\mathbf{s}\cdot \mathbf{E})$ and $\mathbf{m} = 4\pi \gamma_4[\mathbf{s} \times (\boldsymbol{\nabla}\times \mathbf{E})]$ induced by a non-uniform electric field.  The interaction Hamiltonian follows the usual prescription for an induced dipole, e.g., $H_\text{int} = - 2 \pi \gamma_1[\mathbf{s} \times (\boldsymbol{\nabla}\times \mathbf{B})]\cdot \mathbf{E}'$.  Adding these four classical spin-dependent interactions to the spin-independent Hamiltonian  results in a low-energy scattering amplitude consistent with the structure of the LEX of the Compton amplitude in the CoM frame, Eq.~(\ref{amp_com}).

Now we turn to the Majorana condition, recalling that a Majorana fermion is equal to its conjugate under $\mathcal{CPT}$.  Examining the classical Hamiltonian for the $\mathcal{P}$-even, $\mathcal{T}$-even contribution  to Compton scattering, we see that   no fundamental constraints on these polarizabilities arise from the particle's self-conjugate nature. For the spin-independent terms,  it is trivial to show that the Hamiltonian, $H_\text{int} \sim E^2, B^2$, is invariant under $\mathcal{CPT}$.  For the spin-dependent terms, the product of the electric and magnetic fields in $H_\text{int}$ will be even under $\mathcal{CPT}$, and the minus sign that arises from transforming the spin is cancelled by the minus sign that comes from the spatial derivatives.  So there is no {\em a priori} reason for a Majorana fermion's  polarizabilities to vanish for the $\mathcal{P}$-even, $\mathcal{T}$-even contribution to the amplitude.

We now shift to a discussion of the $\mathcal{P}$-odd, $\mathcal{T}$-even portion of the amplitude.  A survey of the expression for the CoM amplitude, Eq.~(\ref{amp_podd_com}), indicates the presence of one spin-independent polarizability and three spin-dependent ones.  As before, we can construct a classical Hamiltonian from these that should reproduce the field theoretic results at low energies.  As above, each term in the classical Hamiltonian will be quadratic in the electric and/or magnetic fields.  By judicious choice (using derivatives and the spin pseudovector), we can construct expressions that are $\mathcal{P}$-odd; however, if we attempt to make each term in the Hamiltonian $\mathcal{C}$-odd (and $\mathcal{T}$-even), we run into a problem.  
Both the electric and magnetic field are odd under charge conjugation, rendering their product even, and the other aforementioned structures (derivatives and spin) are $\mathcal{C}$-even.  With these ingredients, we cannot construct a $\mathcal{C}$-odd term.  We would need an additional $\mathcal{C}$-odd structure (like a current or charge density), but in the case of the Majorana fermion, its self-conjugate nature prohibits such a term from existing.  So, we see at the classical level why $\mathcal{P}$-odd, $\mathcal{T}$-even terms in the Compton amplitude are forbidden for Majorana fermions -- there is no available structure to make interactions $\mathcal{C}$-odd. 

To summarize, for a Dirac spin-$\frac{1}{2}$ fermion there are six independent terms which contribute to the $\mathcal{P}$-even, $\mathcal{T}$-even portion of the Compton scattering amplitude and four terms which contribute to the $\mathcal{P}$-odd, $\mathcal{T}$-even portion of the amplitude.  For a Majorana spin-$\frac{1}{2}$  fermion, the Majorana condition results in no additional constraints on the $\mathcal{P}$-even, $\mathcal{T}$-even portion of the amplitude; however, we find that the $\mathcal{P}$-odd, $\mathcal{T}$-even portion of the amplitude vanishes identically.

\section{Model calculation \label{simp_mod}}

In this section, we use a simple model to compute the anapole moment and structure-dependent polarizabilities for a Majorana fermion, $\chi$,  of mass $m_\chi$. Through explicit calculation, we will see the model-independent results from the previous section borne out in a concrete manner.
The anapole moment is generated through a parity-violating coupling to a scalar $\phi$ of mass $M_\phi$ and fermion $\psi$ of mass $m_f$   The interaction term of the Lagrangian is given by
\begin{equation}
\mathcal{L}_\text{I} =  \overline{\psi}(g_{L}P_L + g_R P_R) \chi  \phi^*+ \mathrm{h.c.} ,  \label{L_int}
\end{equation}
where we define the projections $P_{L,R} = \frac{1}{2}(1 \mp \gamma^5)$.  Assuming $g_L \ne g_R$ (which we take to be real), the chiral projections result in the parity violation necessary to generate an anapole moment.
Other than this interaction, the Dirac fermion and scalar follow the usual rules of QED and scalar electrodynamics; we take their charge to be $e$.  

\begin{figure}
\includegraphics[width=3in]{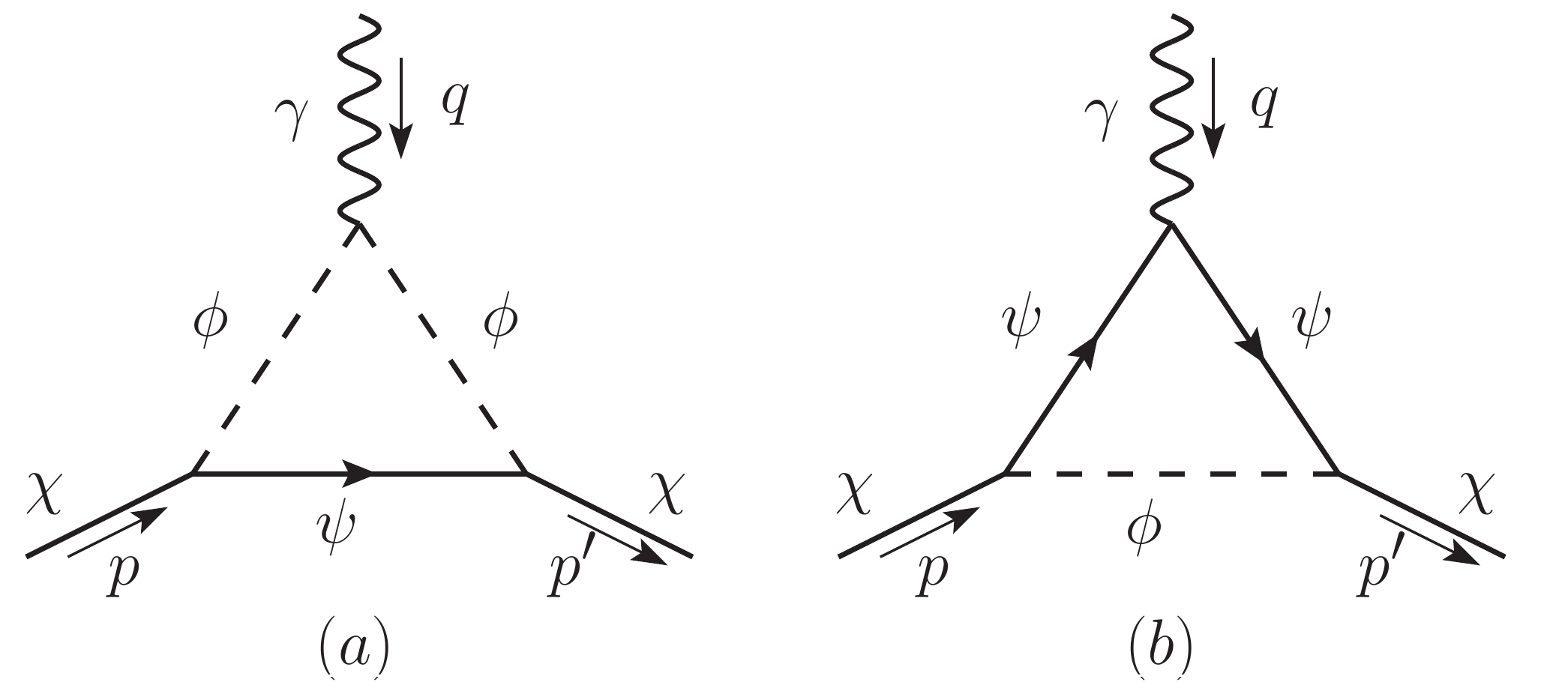}
\caption{ Leading order contributions to the anapole moment for the Majorana fermion. \label{fig_anapole}}
\end{figure}

\subsection{Anapole moment}
The leading order contributions to the anapole moment are depicted in Fig.~\ref{fig_anapole}.  The photon carries momentum $q$ but is not (necessarily) on mass shell whereas both of the Majorana fermion spinors are assumed to be on mass shell so that $\pslash u(p) = m_\chi u(p)$.  Because we are dealing with a Majorana fermion, the usual Feynman rules for Dirac fermions must be adapted; we follow the procedure outlined in Refs.~\cite{denner1,denner2} by Denner, {\it et al.}.  Alternate rules for dealing with Majorana fermions exist, e.g., Ref.~\cite{haber_kane}, and yield equivalent results.  Following Denner, {\it et al.}, we represent fermions in our Feynman diagrams with solid straight lines and distinguish Dirac from Majorana fermions with an arrow.  In addition to the diagrams in Fig.~\ref{fig_anapole},  we include diagrams with reversed fermion flow for the Dirac fermion propagator.   Considering both directions of the fermion flow, the diagrams in panels (a) and (b) of Fig.~\ref{fig_anapole} contribute to the overall amplitude $\mathcal{M}^\mu =  \mathcal{M}^\mu _a+\mathcal{M}^\mu _b$ according to
\begin{eqnarray}
\mathcal{M}^\mu_{a}&=&ie(g_L^2-g_R^2) \int \frac{\mathrm{d}^4 k}{(2\pi)^4}\frac{ (2k^\mu-q^\mu) \bar{u}(p')(\kslash+\pslash)\gamma^5u(p)}{[(k+p)^2-m_f^2][k^2-M_\phi^2][(k-q)^2-M_\phi^2]} \label{Ma}\\
\mathcal{M}^\mu_{b}&=&ie(g_L^2-g_R^2) \int \frac{\mathrm{d}^4 k}{(2\pi)^4}\frac{ \bar{u}(p')[ (\kslash+\qslash)\gamma^\mu\kslash+ m_f^2\gamma^\mu] \gamma^5u(p)}{[(k+q)^2-m_f^2][k^2-m_f^2][(k-p)^2-M_\phi^2]}.  \label{Mb}
\end{eqnarray}
The overall factor of $(g_L^2-g_R^2)$ in the amplitude illustrates the necessity of parity violation in generating the anapole moment.  We recall that the amplitude can be written as  $\mathcal{M}^\mu = \bar{u}(p')  f_A(q^2) \left( q^2 \gamma^\mu - q^\mu \qslash \right) \gamma^5 u(p)$, Eq.~(\ref{vertex}), where the function $f_A(q^2=0)$ represents the static anapole moment.  At low momentum transfer, we find that Eqs.~(\ref{Ma}) and (\ref{Mb}) result in the anapole moment
\begin{equation}
f_A(q^2=0) = \frac{e(g_L^2-g_R^2)}{(8\pi)^2m_\chi^2} \left\{ \left( M_\phi^2 -m_f^2+ \frac{1}{3} m_\chi^2 \right) \int_0^1 \frac{\mathrm{d}x}{m_\chi^2 x^2 +(M_\phi^2 -m_\chi^2 -m_f^2)x +m_f^2} + \log\left( \frac{m_f^2}{M_\phi^2}\right) \right\}.  \label{fA}
\end{equation}
The integral above can be evaluated in closed form (e.g., see the Appendix in Ref.~\cite{dm_n}); however, for simplicity, we will not be concerned with such details and view the anapole vertex as if it were an effective interaction, relevant below some large mass scale.  We suppose that the scalar mass dominates the loop, that is,  $m_f, m_\chi \ll M_\phi$.  Given this, we find
\begin{equation}
f_A(q^2=0) \approx \frac{e(g_L^2-g_R^2)}{(4\pi)^2M_\phi^2} \left[ \frac{1}{3} \log\left( \frac{M_\phi^2}{m_f^2}  \right)- \frac{1}{2} \right].
\end{equation}
As one might anticipate, the form factor is dominated by the inverse square of the heaviest mass in the loop. Additionally, we note that structure of this anapole moment is similar to the computation of the neutrino charge radius in Ref.~\cite{Bernabeu:2000hf}.

\subsection{Polarizabilities}

\begin{figure}
\includegraphics[width=3in]{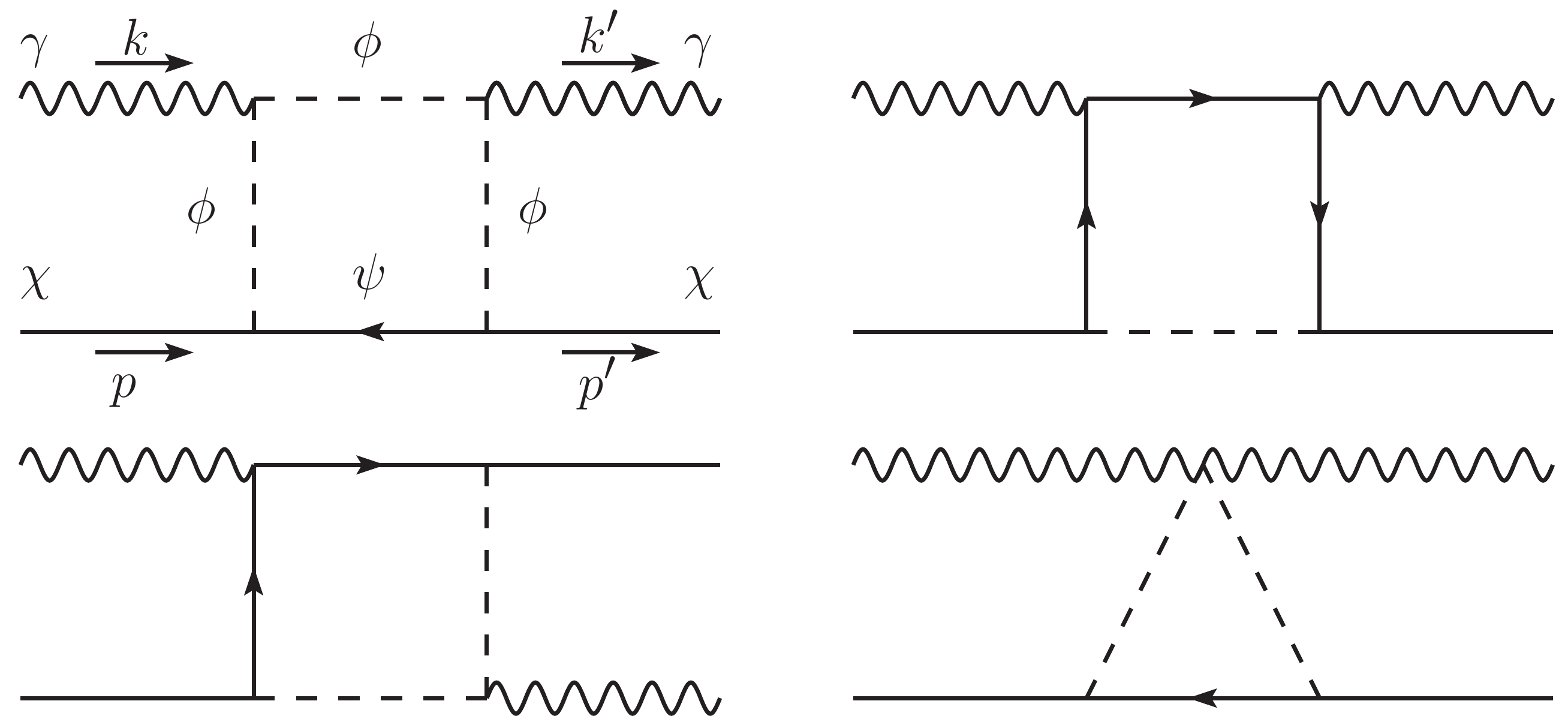}
\caption{ Leading order contributions to Compton scattering via a Majorana fermion. \label{fig_boxes}}
\end{figure}

We now consider a two photon process, computing the Compton amplitude for a Majorana fermion in the CoM frame.     We refer to Fig.~\ref{fig_boxes} for representative Feynman diagrams.   In addition to these, we also include diagrams with reversed fermion flow, and all but the seagull diagram have a partner diagram with ``crossed" photons.   With our simple model, we find that the amplitude conforms to the structure for Majorana fermions discussed in the previous section.  Namely, there are six independent terms in the amplitude with each even under $\mathcal{C}$, $\mathcal{P}$, and $\mathcal{T}$ transformations as in Eq.~(\ref{amp_com}).  There are no terms that are $\mathcal{P}$-odd, $\mathcal{T}$-even, in contrast to the situation that would occur for Dirac fermions.  Finally, we note that if we allowed  complex couplings $g_{L,R} \ne g_{L,R}^*$  (admitting $\mathcal{CP}$-violating interactions)  in Eq.~(\ref{L_int}) then the amplitude would contain $\mathcal{P}$-odd, $\mathcal{T}$-odd terms consistent with those discussed in Ref.~\cite{gorchtein}.

In our computations, we effect a low-energy expansion neglecting terms that are  $\mathcal{O}(\omega^4)$. The exact expressions, to this order, for coefficients $A_1$ through $A_6$ from Eq.~(\ref{amp_com}) are contained in the Appendix.  We confirm that they have the expected structure from the LEX in Eqs.~(\ref{A1_LEX}--\ref{A6_LEX}).  Rather than work with the unwieldy exact expressions, we make some approximations consistent with those made in the previous section.  First, the Compton amplitude is the sum of two sets of terms proportional to the couplings $g_R^2+g_L^2$ and $g_L g_R$.  Here, we will focus only upon the terms proportional to $g_R^2+g_L^2$.  If the interaction in Eq.~(\ref{L_int}) violated parity maximally, say $g_L=0$, then only this first term would survive.  Again, we consider the approximation in which the scalar mass dominates $m_f, m_\chi \ll M_\phi$.  Keeping only the leading order terms in this limit, we find the Majorana fermion's polarizabilities
\al{
\alpha \approx&(g_R^2 + g_L^2 )  \frac{e^2}{(4\pi)^3}  \frac{m_\chi}{M_\phi^4} \Bigg[ \frac{2}{3} \log\left(\frac{M_\phi^2}{m_f^2}\right)-  \frac{5}{6} \Bigg] \label{alpha}, \\
\beta \approx&  (g_R^2 + g_L^2 ) \frac{e^2}{(4\pi)^3} \frac{m_\chi}{M_\phi^4} \Bigg[\frac{2}{3} \log\left(\frac{M_\phi^2}{m_f^2}\right)    - \frac{13}{6}\Bigg], \\
\gamma_1 \approx& -(g_R^2 + g_L^2 ) \frac{e^2}{(4\pi)^3}  \frac{1}{3} 
 \frac{1}{M_\phi^2 m_f^2} \label{g1}, \\
\gamma_2 \approx&   (g_R^2 + g_L^2 ) \frac{e^2}{(4\pi)^3}  \frac{1}{M_\phi^4} \Bigg[     \frac{2}{3} \log\left(\frac{M_\phi^2}{m_f^2}\right) -\frac{3}{2} \Bigg] \label{g2}, \\
\gamma_3 \approx& (g_R^2 + g_L^2 )\frac{e^2}{(4\pi)^3}
        \frac{1}{6}\frac{1}{M_\phi^2m_f^2} \label{g3},\\
\gamma_4 \approx&  -(g_R^2 + g_L^2 ) \frac{e^2}{(4\pi)^3}   \frac{1}{6} \frac{1}{M_\phi^2 m_f^2}.\label{g4}
}
We note that the the spin-independent polarizabilities are commensurate in size particularly for small $m_f$.  Also, the spin-dependent polarizabilities exhibit relationships similar to those computed to leading order for nucleons in chiral perturbation theory; namely, we find $\gamma_3 = -\gamma_4$, $\gamma_2 \approx 0$, and $\gamma_1 = \gamma_4-\gamma_3$ \cite{babusci}.  But, the most interesting feature is the mass dependence of the polarizabilities.

At low energies, the vertex characterizing one- and two-photon interactions with a Majorana fermion can be cast within the framework of an effective field theory.  That is, the anapole moment should capture one-photon interactions and the polarizabilities should capture those for two photons without needing to worry about the details of a UV-complete theory contained in the Feynman diagrams,  Figs.~\ref{fig_anapole} and \ref{fig_boxes}.  
At the Lagrangian level, the anapole interaction takes the form $\mathcal{L}_\text{ana} \sim \overline{\chi}\gamma_\mu \gamma^5 \chi \partial_\nu F^{\mu\nu}$ while the two-photon interaction should be captured by terms $\mathcal{L}_\text{pol} \sim \overline{\chi}( a \mathbbm{1} + b \gamma^5) \chi F^{\mu\nu}F_{\mu\nu}$, cf.~Refs.~\cite{rayleigh_dm,Chen:2013gya, Rajaraman:2012fu}.  The anapole term in the Lagrangian has a mass dimension of 6 so the coupling constant must have mass dimension of negative 2, $g_\text{ana} \sim \frac{1}{\Lambda^2}$ where $\Lambda$ is the energy scale associated with the effective interaction.  For our explicit computation of the anapole moment above, Eq.~(\ref{fA}), we see that this energy scale is set by the (assumed large) scalar mass $\Lambda \sim M_\phi$.  Turning to the two-photon effective Lagrangian, we see that it has a mass dimension of 7, and thus our expectation is that couplings should be characterized by a scale $g_\text{pol} \sim \frac{1}{\Lambda^3} \sim \frac{1}{M_\phi^3}$. Looking at our explicit calculation of the model's polarizabilities, Eqs.~(\ref{alpha}--\ref{g4}), we see that the spin independent polarizabilities roughly meet the expectation $\alpha, \beta \sim \frac{m_\chi}{M_\phi^4}$.  But, three of the spin-dependent polarizabilities have a mass dependence of $\frac{1}{M_\phi^2 m_f^2}$.  On the face of it, processes in which these polarizabilities are relevant are suppressed by only two factors of the scale $\frac{1}{\Lambda}$ rather than three.  This sort of mismatch between EFT expectations and results from a UV-complete theory has been discussed extensively; see Ref.~\cite{DeSimone:2016fbz} and the references therein.

\subsection{Application}

The explicit computation of the anapole moment and polarizabilities for this particular model support the general conclusions of Sec.~\ref{2photon}, but we can also use these polarizabilities  to estimate cross-sections for some low-energy processes.  In particular, we can compute the amplitude for the annihilation of two Majorana fermions into two photons.  This annihilation cross-section is relevant for models of dark matter--both to determine the relic dark-matter density and to inform indirect DM searches.  For these purposes, DM would be non-relativistic, so we compute the annihilation cross-section assuming small relative velocity, $v_\text{rel}\sim 0$;  that is, we would like to estimate the $s$-wave contribution to annihilation, if it exists.

Because real photons do not couple to anapoles, the Born contribution  (with anapole vertex) to the process vanishes, so we move on to the contribution from the box diagrams, Fig.~\ref{fig_boxes}, that  result in the Majorana fermion's polarizabilities.  If the annihilation is $s$-wave, then in the rest frame of the fermion the resulting photons will be emitted back to back with energy $\omega = m_\chi$ and orthogonal polarization,  $\beps' \cdot \beps=0$.    Using crossing symmetry, we can get the amplitude for this process from a computation of the forward Compton amplitude ($k'=k$ and $p'=p$)
in the scatterer's rest frame, $p=(m_\chi, \mathbf{0})$, with photon momentum $k =\omega (1,  \khat)$ where $\omega= m_\chi$.  Boosting to the CoM frame, we  can use the decomposition of the Compton amplitude in Eq.~(\ref{amp_com}) to see which terms would contribute to the process. Only one term survives, $\mathcal{M} = i A_3 \xi'^\dagger [\bsig \cdot (\beps \times \beps')]\xi$, and this term retains its structure in the fermion rest frame.  In the CoM frame, we recall the LEX for this term $A_3 \approx 
-8\pi m_\chi [ \gamma_1 -(\gamma_2 + 2 \gamma_4) \khat' \cdot \khat] \tilde{\omega}^3$ where $\tilde{\omega}$ is the boosted photon energy ($m_\chi/\sqrt{3}$).  Boosting this back to the fermion rest frame, the leading order term remains, namely $A_3 \approx 
-8\pi m_\chi [ \gamma_1 -(\gamma_2 + 2 \gamma_4) \khat' \cdot \khat] {\omega}^3$, but errors that are $\mathcal{O}(1)$ accrue.  Still, this is sufficient to obtain an order of magnitude estimate. Using crossing symmetry and  the polarizabilities in Eq.~(\ref{g1},\ref{g2},\ref{g4}), we estimate the $s$-wave annihilation  amplitude to be $\mathcal{M}(\chi\chi \to \gamma \gamma) \approx i  \frac{4 }{3} g_R^2 \alpha
 \frac{ m_\chi^4}{M_\phi^2 m_f^2} \xi'^\dagger[\bsig \cdot (\beps \times \beps')] \xi$, assuming $g_L \equiv 0$.   Averaging over spins and summing over final polarization states, we find 
 \begin{equation}
 |v_\text{rel}| \frac{\mathrm{d} }{\mathrm{d} \Omega} \sigma_{\chi\chi\to\gamma\gamma} \approx \frac{g_R^4 \alpha^2}{(6\pi)^2} 
 \frac{ m_\chi^6}{M_\phi^4 m_f^4}.
\end{equation}
Per our discussion on EFT, we would na\"ively expect this cross-section to be suppressed by at least a factor of $M_\phi^{-6}$, but we see that our estimate from the explicit calculation is much larger.   Given this and the fact that  the annihilation is $s$-wave, it could have a significant impact upon determining the relic density in a theory of Majorana DM.

\section{Conclusions}

Because the electromagnetic properties of Majorana fermions are severely constrained, these fermions cannot couple to a single real photon, but if we move beyond the Born contribution, Majorana fermions can interact with real photons in a two-photon process.   For a Majorana scatterer, we have shown  generally that contributions to the Compton amplitude are not necessarily forbidden as long as the process is separately invariant under the discrete symmetries $\mathcal{C}$, $\mathcal{P}$, and $\mathcal{T}$.  However, there are some restrictions upon the Compton amplitude; namely, contributions to the amplitude that are $\mathcal{P}$-odd and $\mathcal{T}$-even must vanish because Majorana fermions are self-conjugate fields.  These general findings were borne out in an explicit computation of the polarizabilities of a Majorana fermion assuming a simple model.  From the explicit computation, we learned that some of the spin-dependent polarizabilities were not suppressed by the appropriate mass scale expected in an effective field theory.  Of consequence is the fact that Majorana fermions can undergo $s$-wave annihilation into two photons with a much greater cross-section than one might na\"ively expect.

\section{Appendix}

We include the full expression for the CoM Compton amplitude coefficients $A_j$ for the simple model discussed in Sec.~\ref{simp_mod} accurate to $\mathcal{O}(\omega^3)$.  For the portion of the amplitude that is $\mathcal{P}$-even and $\mathcal{T}$-even, the amplitude is the sum of terms proportional to the factor $(g_R^2+g_L^2)$  and  $g_Rg_L$ which we will denote as $A_j^S$ and $A_j^D$ respectively so that $A_j = A^S_j + A^D_j$.  We find for $A^S_j$ 
\al{
A^S_1 =&      -  (g_R^2 + g_L^2 )  \frac{e^2}{(4\pi)^2 } \int_0^1 \mathrm{d}x \, \Bigg\{\omega^2 \Bigg[  
\frac{m_\chi^2}{P(x)^2}\Bigg( - \frac{4}{3}  x + \frac{2}{3} x^3  \Bigg)
 \nonumber\\
&+ \frac{m_\chi^4}{P(x)^3} \Bigg(- \frac{8}{3} x^3 + \frac{16}{3} x^4 - \frac{8}{3} x^5 \Bigg)+\cos \theta \frac{m_\chi^2}{P(x)^2} \Bigg( - \frac{4}{3} x + 4 x^2 -   2x^3 \Bigg) \Bigg]\nonumber \\
&+ \omega^3 \Bigg[  \frac{m_\chi}{P(x)^2} \Bigg(  -\frac{8}{3}  x + 4 x^2 - \frac{4}{3}  x^3 \Bigg)
+\frac{m_\chi^3 }{P(x)^3}\Bigg(- \frac{8}{3} x^3 + \frac{16}{3} x^4 - \frac{8}{3}  x^5  \Bigg) \Bigg](1+\cos \theta)     \Bigg\} +\mathcal{O}(\omega^4),
\\
 A^S_2 =& - (g_R^2 + g_L^2 )  \frac{e^2}{(4\pi)^2 }    \int_0^1 \mathrm{d}x\,  \Bigg\{\omega^2  \frac{m_\chi^2}{P(x)^2} \Bigg( \frac{4}{3} x  -4 x^2 +2x^3 \Bigg) \nonumber \\
 & + \omega^3 \Bigg[  \frac{m_\chi}{P(x)^2} \Bigg( \frac{8}{3} x - 4  x^2 + \frac{4}{3} x^3  \Bigg)
 + \frac{m_\chi^3}{P(x)^3} \Bigg(\frac{8}{3}   x^3- \frac{16}{3}    x^4 + \frac{8}{3}   x^5 \Bigg) \Bigg] \Bigg\}+\mathcal{O}(\omega^4),
\\
A^S_3 =&  -  (g_R^2 + g_L^2 )\frac{e^2}{(4\pi)^2 }    \int_0^1 \mathrm{d}x\,  \omega^3 \Bigg\{ \frac{m_\chi}{P(x)^2}\Bigg( - \frac{2}{3}  + 2 x^2 - \frac{4}{3} x^3\Bigg) \nonumber\\
& +\cos \theta \Bigg[\frac{m_\chi}{P(x)^2} \Bigg( \frac{2}{3} - 2 x^2 + \frac{4}{3} x^3 \Bigg) 
+ \frac{m_\chi^3}{P(x)^3} \Bigg( \frac{8}{3} x^2 - 8 x^3 + 8 x^4 - \frac{8}{3} x^5\Bigg) \Bigg] \Bigg\}+\mathcal{O}(\omega^4),
\\
A^S_4 =&  -(g_R^2 + g_L^2 ) \frac{e^2}{(4\pi)^2 }   \int_0^1 \mathrm{d}x \, \omega^3 \Bigg[
       \frac{m_\chi}{P(x)^2} \Bigg ( \frac{4}{3}  x -  2x^2 + \frac{2}{3}  x^3 \Bigg)
       +\frac{m_\chi^3}{P(x)^3} \Bigg( \frac{8}{3} x^2 - 8 x^3 + 8 x^4 - \frac{8}{3} x^5 \Bigg)\Bigg]+\mathcal{O}(\omega^4),
\\
A^S_5 =& -  (g_R^2 + g_L^2 ) \frac{e^2}{(4\pi)^2 }  \int_0^1 \mathrm{d}x \, \omega^3 \Bigg[
       \frac{m_\chi}{P(x)^2}  \Bigg( \frac{1}{3} + \frac{2}{3} x - 2 x^2 +  x^3 \Bigg)
       + \frac{m_\chi^3}{P(x)^3}  \Bigg( \frac{8}{3} x^2 - 8x^3 + 8 x^4 - \frac{8}{3} x^5 \Bigg)\Bigg]+\mathcal{O}(\omega^4),
\\
A^S_6=&- (g_R^2 + g_L^2 ) \frac{e^2}{(4\pi)^2 }  \int_0^1 \mathrm{d}x \, \omega^3 \Bigg[
       \frac{m_\chi}{P(x)^2}  \Bigg( - \frac{1}{3}  -\frac{ 2}{3}x + 2 x^2 - x^3\Bigg) 
        + \frac{m_\chi^3}{P(x)^3} \Bigg( - \frac{4}{3} x^2 + 4 x^3 - 4 x^4 + \frac{4}{3} x^5 \Bigg) \Bigg] +\mathcal{O}(\omega^4),
} 
where $P(x):= m_\chi^2 x^2 +(M_\phi^2 - m_\chi^2 -m_f^2) x + m_f^2$.
The final terms are
\al{
  A_1^D =&
    -2g_R g_L     \frac{e^2}{(4\pi)^2 } \int_0^1 \mathrm{d}x  \Bigg\{\omega^2 \Bigg[  
\frac{m_\chi m_f}{P_1^2}\Bigg( \frac{4}{3} - 4   x + 2x^2   \Bigg)
 + \frac{m_\chi^3 m_f}{P_1^3} \Bigg( - \frac{8}{3} x^2 + \frac{16}{3} 
x^3 - \frac{8}{3}x^4  \Bigg)
 \nonumber\\
&+\cos \theta \frac{m_\chi m_f }{P_1^2} \Bigg(  - \frac{4}{3} + 4 x -  2x^2  \Bigg) \Bigg]+ \omega^3 \Bigg[ 
\frac{ m_\chi^2 m_f }{P_1^3}\Bigg(- \frac{8}{3} x^2 + \frac{16}{3} x^3 - \frac{8}{3} x^4   \Bigg) \Bigg] (1+\cos \theta)    \Bigg\}   +\mathcal{O}(\omega^4),\\
   A_2^D =&
 - 2g_R g_L    \frac{e^2}{(4\pi)^2 } \int_0^1 \mathrm{d}x  \Bigg\{\omega^2    
\frac{ m_\chi m_f }{P_1^2}\Bigg(    \frac{4}{3}   - 4  x + 2 x^2  \Bigg) + \omega^3
\frac{ m_\chi^2 m_f }{P_1^3}\Bigg( \frac{8}{3} x^2 - \frac{16}{3}x^3 + \frac{8}{3}x^4 \Bigg)     \Bigg\}   +\mathcal{O}(\omega^4),\\
A_3^D =& -2g_R g_L\frac{e^2}{(4\pi)^2 } \int_0^1 \mathrm{d}x  \omega^3 \Bigg\{  
\frac{m_f}{P_1^2} (  - 2 + 4 x - 2 x^2 ) +\frac{m_f m_\chi^2}{P_1^3}\Bigg(  -\frac{8}{3}x + 8x^2 - 8 x^3 +\frac{ 8}{3}x^4 \Bigg)\nonumber \\
&+\cos \theta \Bigg[ 
\frac{m_f}{P_1^2} (2 - 4  x + 2  x^2) +\frac{m_f m_\chi^2}{P_1^3}\Bigg(  \frac{8}{3}x - 8x^2 + 8 x^3 -\frac{ 8}{3}x^4 \Bigg) \Bigg]
\Bigg\}+\mathcal{O}(\omega^4),\\
A_4^D = &  -  2g_R g_L\frac{e^2}{(4\pi)^2 } \int_0^1 \mathrm{d}x  \omega^3     \frac{m_\chi^2 m_f}{P_1^3} \Bigg( \frac{8}{3} x - 8 x^2 + 8 x^3 - \frac{8}{3} x^4 \Bigg) +\mathcal{O}(\omega^4),
\\
A_5^D = &   - 2g_R g_L\frac{e^2}{(4\pi)^2 } \int_0^1 \mathrm{d}x  \omega^3      \Bigg\{  \frac{m_f}{P_1^2} \Bigg(  1 - 2 x + 
         x^2 \Bigg)
+\frac{m_\chi^2 m_f}{P_1^2} \Bigg(  \frac{8}{3} x - 8  x^2 + 8  x^3 - \frac{8}{3}x^4 \Bigg) +\mathcal{O}(\omega^4),
\\
 A^D_6 =& -2g_R g_L \frac{e^2}{(4\pi)^2 } \int_0^1 \mathrm{d}x  \omega^3  
      \Bigg\{ \frac{m_f}{P_1^2} (  -  1  + 2 x
          - x^2 )
       + \frac{m_\chi^2 m_f}{P_1^3}\Bigg(  - \frac{4}{3} x + 4x^2 - 4x^3 + \frac{4}{3}x^4 \Bigg) \Bigg\}+\mathcal{O}(\omega^4).
}

\section{ACKNOWLEDGMENTS}
This work was funded, in part, by a Mellon Junior Sabbatical Fellowship  from the University of Puget Sound.

\bibliography{biblio}

\end{document}